\documentclass[twocolumn]{revtex4-2}
\usepackage{environ}
\NewEnviron{eqsp}{%
  \begin{equation}
    \begin{split}
      \BODY
    \end{split}
  \end{equation}
}
\def\CA{{\cal A}}

\def\CH {{\cal H}}
\def\CI {{\cal I}}

\def\CN {{\cal N}}
\def\CO {{\cal O}}

\def\CO {{\cal O}}

\def\CH {{\cal H}}
\def\CI {{{\cal I}}}

\usepackage{amsmath, amssymb, slashed, epsf, color, graphicx, latexsym, bm, amsfonts}
\newtheorem{thm}{Theorem}[section]
\newtheorem{prop}{Proposition}[section]
\newtheorem{lemma}{Lemma}[section]

\usepackage{graphicx}
\usepackage{dcolumn}
\usepackage{bm}

\begin{document}

\preprint{}

\title{Macdonald Index From Refined Kontsevich-Soibelman  Operator}

\author{George Andrews}
\email{gea1@psu.edu}
 \affiliation{Department of Mathematics, Penn State University, University Park, PA 16802}
\author{Anindya Banerjee}%
 \email{ab1702@scarletmail.rutgers.edu}
\affiliation{%
 Department of Physics, University of Cincinnati, Cincinnati, OH 45211
}%

\author{Ranveer Kumar Singh}
\email{rks158@scarletmail.rutgers.edu}
\affiliation{
NHETC and Department of Physics and Astronomy, Rutgers University, 126 Frelinghuysen Rd., Piscataway NJ 08855, USA
}%
\author{Runkai Tao}
\email{runkai.tao@physics.rutgers.edu}
\affiliation{%
NHETC and Department of Physics and Astronomy, Rutgers University, 126 Frelinghuysen Rd., Piscataway NJ 08855, USA
}%

\date{\today}

\begin{abstract}
We propose a refinement of the Kontsevich-Soibelman  operator for a class of 4d $\mathcal{N}=2$ superconformal field theories characterized by the following conditions: (1) their Coulomb branch admits a source/sink chamber, i.e., a chamber in which the BPS quiver consists of only source and sink nodes, (2) The nodes with valency greater than 2 of the BPS quiver in a source/sink chamber are either all sources or all sinks. We present strong evidence that the trace of this refined  operator is related to the Macdonald index of the theory. In particular, we conjecture new closed form expressions for the Macdonald indices of the $(A_1,\mathfrak{g})$ Argyres-Douglas theories for any simply-laced Lie algebra $\mathfrak{g}$.   
\end{abstract}

\maketitle


\section{\label{sec:intro}Introduction}
All known examples of interacting 4d $\mathcal{N}=2$ quantum field theories have a continuous moduli space of vacua known as the Coulomb branch. This is a singular manifold, and powerful techniques for analyzing the (protected or BPS) spectrum and dynamics of these theories may be derived from the operation of moving on this moduli space \cite{Seiberg:1994rs,Seiberg:1994aj}. In particular, many observables are related to representations of the monodromy group associated with the singularities. 

In this work, we will make use of such an object, the $q$-deformed version \footnote{The original Kontsevich-Soibelman operator was introduced in context of Donaldson-Thomas invariants in \cite{Kontsevich:2008fj} and used in context of 4d $\CN=2$ theories in \cite{Gaiotto:2010okc} to study wall crossing.} of the Kontsevich-Soibelman monodromy operator $M(q)$ \cite{Dimofte:2009bv,Dimofte:2009tm}, to understand the spectrum of BPS states in these theories \cite{Gaiotto:2010okc}. In particular, we will focus on superconformal field theories, where the full $\text{SU}(2)\times \mathrm{U}(1)$ symmetry is non-anomalous and by virtue of the state-operator correspondence, one may study BPS local operators. Furthermore, in such SCFTs, the monodromy operator satisfies $M^r=I$ for some finite positive integer $r$ \cite{Cecotti:2010fi,Cecotti:2015lab}. We will make use of a counting function that tracks the multiplicities of various superconformal representations that appear in the spectrum of local operators in the theory (note that the superconformal index has ambiguities associated with degeneracies and mutually cancelling contributions). The superconformal index becomes more unambiguous when enumerating the operator content for more protected local operators (like half-BPS operators). We will focus here on quarter-BPS operators where there are some ambiguities, so that these lie at the present threshold of our ability to reconstruct from the index. 

The SCFT lives at the origin of moduli space, and moving out on the moduli space is accomplished by turning on vacuum expectation values for certain universal scalar operators that exist in these theories. The underlying argument in our work is that the low energy effective theory obtained in the infrared by moving away from the origin of the moduli space contains enough information to partially reconstruct the BPS spectrum of the SCFT living at the origin. Such a use for the monodromy operator was first advocated in \cite{Cordova:2015nma} where the Schur indices of several classes of 4d $\mathcal{N}=2$ SCFTs was computed. In this letter, we propose a prescription to generalize the monodromy operator for a ``special'' class of 4d $\mathcal{N}=2$ SCFTs. We will use this generalized operator to study the same Schur subsector of the local operator spectrum of these theories \cite{Beem:2013sza}, but compute a refined counting function associated with this sector known as the Macdonald index \cite{Gadde:2011uv}. This may also be understood as a special limit of the full superconformal index. See \cite{Buican:2015tda,Bhargava:2023hsc,Song:2016yfd,Andrews:2025krn,Kang:2025zub} for other existing ways of computing the Macdonald index of 4d $\CN=2$ theories. We expect a similar generalization to work for a much larger class of SCFTs. In the present work, we use our prescription to conjecture new closed form expression for the Macdonald indices of $(A_1,\mathfrak{g})$ Argyres-Douglas theories. 
\\\\
\textit{Note Added.} After we submitted our paper to the arxiv, the paper \cite{Kim:2025klh} appeared which also conjectures expressions for the Macdonald indices of $(A_1,\mathfrak{g})$ theories using the KS monodromy operator. 
\section{Kontsevich-Soibelman Operator}
Any 4d $\CN=2$ theory at a generic point on the Coulomb branch is IR free and is characterized by a $\mathrm{U(1)}^r$ gauge theory. The massive spectrum of the theory consists of particles carrying electric and magnetic charges valued in the charge lattice $\Gamma$ equipped with a Dirac pairing $\langle\cdot,\cdot\rangle$. The spectrum of BPS states is particularly interesting because they behave nicely under certain supersymmetry-preserving RG flows. The spectrum of BPS states with a given charge $\gamma\in\Gamma$ is captured by the protected spin character (PSC) defined as follows \cite{Gaiotto:2010be}: let $\CH^{\gamma}_{\mathrm{int}}$ be the internal Hilbert space of the BPS particle with charge $\gamma$. Then the PSC of $\gamma$ is defined by 
\begin{eqsp}
\Omega(\gamma,y):=\mathrm{Tr}_{\CH_{\text{int}}^\gamma}(-1)^{R}y^{J+R}=\sum_{n\in\mathbb{Z}}\Omega_n(\gamma)y^n~,
\end{eqsp}
where $J$ is the generator of the Cartan subalgebra of the little group SU(2) of the Lorentz group SO(1,3) and $R$ is the Cartan generator of the SU(2) R-symmetry group. As one moves in the moduli space, the BPS spectrum is piecewise constant but may jump discontinuously across certain codimension one subloci, called walls of marginal stability. To understand this jump, one needs to construct wall-crossing invariants. Define the quantum torus algebra to be the formal span of generators 
\begin{eqsp}
    \CA:=\mathrm{Span}_{\mathbb{C}}\{X_\gamma:\gamma\in\Gamma\}~,
\end{eqsp}
with the multiplication on $\CA$ defined as
\begin{eqsp}\label{eq:quant_tor_alg}
    X_\gamma X_{\gamma^{\prime}}=q^{\frac{\left\langle\gamma, \gamma^{\prime}\right\rangle}{2}} X_{\gamma+\gamma^{\prime}}=q^{\left\langle\gamma, \gamma^{\prime}\right\rangle} X_{\gamma^{\prime}} X_\gamma~,
\end{eqsp}
where $q$ is a formal variable. For each $\gamma\in\Gamma$, we construct the following element of $\CA$:
\begin{eqsp}
    U_\gamma=\prod_{n \in \mathbb{Z}} E_q\left((-1)^n q^{n / 2} X_\gamma\right)^{(-1)^n \Omega_n(\gamma)}~,
\end{eqsp}
where $E_q(X)$ is the quantum dilogarithm
\begin{eqsp}
    E_q(z):=\prod_{i=0}^{\infty}\left(1+q^{i+\frac{1}{2}} z\right)^{-1}=\sum_{n=0}^{\infty} \frac{\left(-q^{\frac{1}{2}} z\right)^n}{(q)_n}~.
\end{eqsp}
The Kontsevich-Soibelman (KS) operator is defined as 
\begin{eqsp}
    \CO(q):=\prod_{\gamma \in \Gamma}^{\curvearrowright} U_\gamma~,
\end{eqsp}
where the product is taken in the order of increasing $\text{Arg}(Z(\gamma))$ with respect to an arbitrarily chosen phase $\theta$, where $Z(\gamma)$ is the central charge corresponding to the charge $\gamma$. This operator is related to the KS operator by $\CO(q)=M(q)^{-1}$. The celebrated result of Kontsevich and Soibelman \cite{Kontsevich:2008fj} is that the operator $\CO(q)$ is a wall crossing invariant. Thus, if one can relate physically relevant quantities to this operator, then one can compute it at any point in the moduli space and glean information about the UV theory at the origin of the moduli space. This was the motivation behind the conjecture of Cordova-Shao \cite{Cordova:2015nma} relating the Schur index of a 4d $\CN=2$ SCFT to the trace of the KS operator.  
\subsection{Trace Of KS Operator And Schur Index}
Let us begin by quoting the conjecture from \cite{Cordova:2015nma}: it states that the Schur index of a 4d $\CN=2$ SCFT is given by an appropriately defined trace: 
\begin{eqsp}\label{eq:CS_conj}
    \CI_{\mathrm{S}}(q,z_1,\dots,z_{N_f}):=(q)_\infty^{2r}\,\mathrm{Tr}\,\CO(q)~,
\end{eqsp}
where $r$ is the rank of the theory, $z_1,\dots,z_{N_f}$ are fugacities for flavor symmetries and 
\begin{eqsp}
    (a)_\infty:=(a;q)_\infty:=\prod_{i=0}^\infty(1-aq^i)~.
\end{eqsp}
The factor $(q)_\infty^{2r}$ is simply the Schur index of $r$ free U(1) vector multiplets which is the contribution of the free $\text{U(1)}^r$ vector multiplets at a generic point on the Coulomb branch. 
Recall that a charge $\gamma \in\Gamma$ is called a flavor charge if $\langle\gamma,\Gamma\rangle=0$. Let $\{\gamma_{f_i}\}_{i=1}^{N_f}$ be an integral basis of the sublattice of flavor charges. For any charge $\gamma$, let 
\begin{eqsp}
    \gamma_f=\sum_{i=1}^{N_f}f_i(\gamma)\gamma_{f_i}~,
\end{eqsp}
be the projection on to the subspace of flavor charges. 
We define 
\begin{eqsp}\label{eq:trace_X}
    \mathrm{Tr}[X_\gamma]=\begin{cases}\prod_i \operatorname{Tr}\left[X_{\gamma_{f_i}}\right]^{f_i(\gamma)} & \left\langle\gamma, \Gamma\right\rangle=0~, \\ 0 & \text { else },\end{cases}~.
\end{eqsp}
Imposing the cyclicity of the trace, the trace of any element of $\CA$ can then be written in terms of $q$ and $N_f$ variables $\text{Tr}[X_{\gamma_{f_i}}]$. These variables are functions of the flavor fugacities $z_1,\dots,z_n$ and the explicit functional relation can be determined as explained in \cite{Cordova:2015nma}. In \cite{Cordova:2015nma}, the trace of the KS operator was computed for a large class of examples and the conjecture \eqref{eq:CS_conj} was verified. 
\section{The Refined KS Operator}
In this section, we define a refined KS operator for a ``special'' class of 4d $\CN=2$ SCFTs. 
\subsection{BPS Quiver And Source/Sink Chamber}
The BPS quiver of a 4d $\CN=2$ theory \cite{Denef:2002ru} captures the spectrum of BPS states of the theory at a given point the Coulomb branch. We will not go into relevance of quivers in 4d $\mathcal{N}=2$ theories in this work, and instead refer readers to \cite{Denef:2002ru, Denef:2007vg, Manschot:2010qz, Manschot:2012rx} for more details. The rules to draw the BPS quiver are as follows \cite{Cecotti:2010fi,Alim:2011ae}:
\begin{enumerate}
    \item Draw a node for each basis element of the charge lattice. 
    \item If for two nodes $\gamma_i,\gamma_j$, $\langle\gamma_i,\gamma_j\rangle>0$, draw $\langle\gamma_i,\gamma_j\rangle$ arrows from node $i$ to node $j$ and if $\langle\gamma_i,\gamma_j\rangle<0$ then draw $|\langle\gamma_i,\gamma_j\rangle|$ arrows from $j$ to $i$.
\end{enumerate}
Note that the quiver depends on a choice of basis and choosing a different basis gives a different quiver.  
Some 4d $\CN=2$ theories admit a source/sink chamber, meaning that the charge lattice at a given point in this chamber admits a basis in which the BPS quiver has only source and sink nodes \footnote{A node with all arrows emanating from it is called a source node and a node with all arrows ending on it is called a sink node.}.  
The (conjectured) crucial property \cite{Cecotti:2010fi,Alim:2011ae} of the source/sink chamber is that the central charges corresponding to the nodes satisfy
\begin{eqsp}
    \mathrm{Arg}(Z(\gamma))>\mathrm{Arg}(Z(\gamma'))~,~~ \text{for all}~\gamma\in\text{sink},~\gamma'\in\mathrm{source}~.
\end{eqsp}
Moreover, it is clear that 
\begin{eqsp}
[X_\gamma, X_{\gamma'}]=0~,\quad \gamma,\gamma'\in\text{sources, or},~~\gamma,\gamma'\in\text{sinks}~.\end{eqsp}
Thus, one can compute the KS operator in the source/sink chamber:
\begin{eqsp}
    \CO(q)=\prod_{\gamma\in\text{source}}U_{-\gamma}\prod_{\gamma\in\text{sink}}U_{-\gamma}\prod_{\gamma\in\text{source}}U_{\gamma}\prod_{\gamma\in\text{sink}}U_{\gamma}~.
\end{eqsp}
\subsection{Refined KS Operator}
\begin{figure}[h]
    \centering
    \includegraphics[width=1\linewidth]{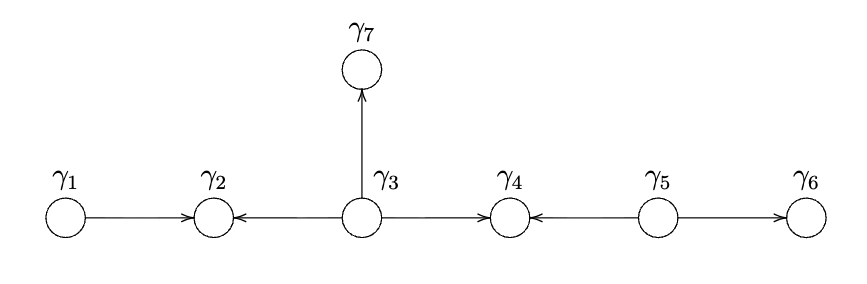}
    \caption{BPS quiver of the $(A_1,E_7)$ theory in the source/sink chamber. Figure adapted from \cite{Cordova:2015nma}.}
    \label{fig:DE7}
\end{figure}

We consider 4d $\CN=2$ SCFTs which satisfy the following two properties:
\begin{enumerate}
    \item The theory admits a source/sink chamber.
    \item The nodes in the BPS quiver at a point in the source/sink chamber with valency greater than 2 are either all sources or all sinks. 
\end{enumerate}
The $(A_1,\mathfrak{g})$ Argyres-Douglas theories introduced in \cite{Cecotti:2010fi} are examples of this special class of theories. Their BPS quivers are simply-laced tree graphs that have a finite number of mutation-equivalent graphs in their equivalence class under quiver mutations. See \cite[Footnote 10]{Alim:2011ae} for more details. Moreover, their moduli space admits a source/sink chamber. Each node in the BPS quiver in the source/sink chamber represents a single hypermultiplet with trivial internal degrees of freedom. Moreover, the quivers for these theories shown in Figure \ref{fig:A2n}--\ref{fig:DE7} clearly satisfy the second condition. 
Define the functions  
    \begin{eqsp}\label{eq:Eqt_def}
        E_{q,T}(X_\gamma)&=\sum_{n=0}^\infty\frac{(-(qT)^{\frac{1}{2}}X_\gamma)^n}{(q)_n},\\\widetilde{E}_{q,T}(X_\gamma)&=\sum_{n=0}^\infty\frac{(-q^{\frac{1}{2}}X_\gamma)^n}{(qT)_n}~.
    \end{eqsp}
Define the refined KS operator in the source/sink chamber by
\begin{eqsp}
\CO(q,T):=&\prod_{\gamma\in\text{source}}\prod_{n\in\mathbb{Z}}E_{q,T}((-1)^nq^{n/2}X_{-\gamma})^{(-1)^n\Omega_n(-\gamma)}\\\times&\prod_{\gamma\in\text{sink}}\prod_{n\in\mathbb{Z}}\widetilde{E}_{q,T}((-1)^nq^{n/2}X_{-\gamma})^{(-1)^n\Omega_n(-\gamma)}\\\times&\prod_{\gamma\in\text{source}}\prod_{n\in\mathbb{Z}}E_{q,T}((-1)^nq^{n/2}X_{\gamma})^{(-1)^n\Omega_n(\gamma)}\\\times &\prod_{\gamma\in\text{sink}}\prod_{n\in\mathbb{Z}}E_{q}((-1)^nq^{n/2}X_{\gamma})^{(-1)^n\Omega_n(\gamma)}~.
\end{eqsp}
\\\\
If the BPS quiver has a source node with valency greater than 2, then we define the refined KS operator as:
\begin{eqsp}\label{eq:KS_source>2}
\CO(q,T):=&\prod_{\gamma\in\text{source}}\prod_{n\in\mathbb{Z}}\widetilde{E}_{q,T}((-1)^nq^{n/2}X_{-\gamma})^{(-1)^n\Omega_n(-\gamma)}\\\times &\prod_{\gamma\in\text{sink}}\prod_{n\in\mathbb{Z}}E_{q,T}((-1)^nq^{n/2}X_{-\gamma})^{(-1)^n\Omega_n(-\gamma)}\\\times&\prod_{\gamma\in\text{source}}\prod_{n\in\mathbb{Z}}E_{q}((-1)^nq^{n/2}X_{\gamma})^{(-1)^n\Omega_n(\gamma)}\\\times &\prod_{\gamma\in\text{sink}}\prod_{n\in\mathbb{Z}}E_{q,T}((-1)^nq^{n/2}X_{\gamma})^{(-1)^n\Omega_n(\gamma)}~.
\end{eqsp}
Note that, at the moment, we can define the refined KS operator only in the source/sink chamber. 
\section{Macdonald Index From Modified KS Operator: A Conjecture}
Recall that the Macdonald index of the free vector multiplet is given by 
\begin{eqsp}
    \CI_{\text{M}}^{\text{vec}}(q,T)=(q)_\infty(qT)_\infty~.
\end{eqsp}
We conjecture that the Macdonald index of the class of 4d $\CN=2$ SCFTs for which the modified KS operator is defined is given by
\begin{eqsp}
    \CI_{\text{M}}(q,T,z_1,\dots,z_{N_f})=(q)^r_\infty(qT)^r_\infty\text{Tr}\,\CO(q,T)~,
\end{eqsp}
where $r$ is the rank of the theory. 
Assuming the conjecture in \cite{Cordova:2015nma}, this formula trivially satisfies the correct Schur limit $T\to 1$. 
\subsection{Evidence For the Conjecture}
We now use our conjecture to calculate the Macdonald index of $(A_1,\mathfrak{g})$ Argyres-Douglas theories. In this section, we present the detailed computation for the $(A_1,A_k)$ theory and summarize our conjectures for other $(A_1,\mathfrak{g})$ theories. 
\subsubsection{$(A_1,A_{k})$ Theory}
\begin{figure}[h]
    \centering
    \includegraphics[width=1\linewidth]{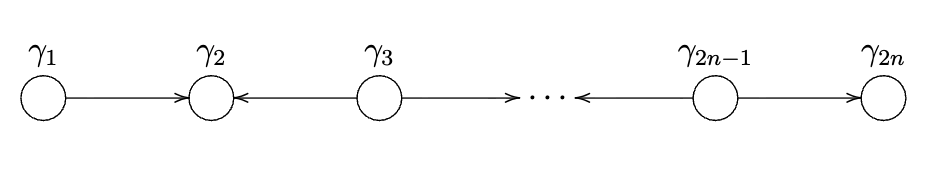}
    \caption{BPS quiver of the $(A_1,A_{2n})$ theory in the source/sink chamber. Figure adapted from \cite{Cordova:2015nma}.}
    \label{fig:A2n}
\end{figure}

Let us start with the $k=2n$ case. The BPS quiver of the $(A_1,A_{2n})$ theory in the source/sink chamber is shown in Figure \ref{fig:A2n}.

Each node in the quiver represents a single hypermultiplet. The refined KS operator is thus given by
\begin{figure}[h]
    \centering
    \includegraphics[width=1\linewidth]{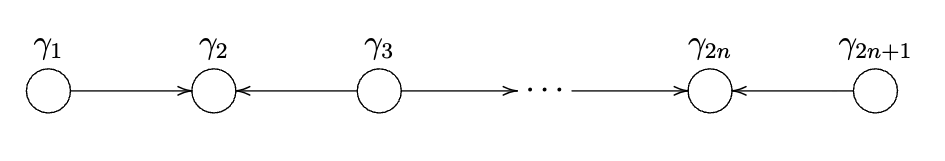}
    \caption{BPS quiver of the $(A_1,A_{2n+1})$ theory in the source/sink chamber. Figure adapted from \cite{Cordova:2015nma}.}
    \label{fig:A2n+1}
\end{figure}
\begin{eqsp}
\CO(q,T)&=\prod_{i=1}^{n}E_{q,T}(X_{-\gamma_{2i-1}})\prod_{i=1}^n\widetilde{E}_{q,T}(X_{-\gamma_{2i}})\\\times&\prod_{i=1}^{n}E_{q,T}(X_{\gamma_{2i-1}})\prod_{i=1}^nE_{q}(X_{\gamma_{2i}})~.
\end{eqsp}

Using the formula for the functions $E_{q,T}$ and $\widetilde{E}_{q,T}$, we get 
\begin{widetext}
\begin{eqsp}
\CO(q,T)=\sum_{\substack{\ell_1, \ldots, \ell_{2 n} \geq 0 \\
k_1, \ldots, k_{2 n} \geq 0}} \frac{(-1)^{\sum_{i=1}^{2 n}\left(k_i+\ell_i\right)} q^{\frac{1}{2} \sum_{i=1}^{2 n}\left(k_{i}+ \ell_i\right)}T^{\frac{1}{2}\sum_{i=1}^n(\ell_{2i-1}+k_{2i-1})}}{\prod_{i=1}^n(q)_{k_{2 i-1}}(q)_{\ell_{2 i-1}}\prod_{i=1}^n(q)_{k_{2i}}(q T)_{\ell_{2 i}} } \prod_{i=1}^n X_{-\gamma_{2 i-1}}^{\ell_{2 i-1}}\prod_{i=1}^n X_{-\gamma_{2 i}}^{\ell_{2 i}} \prod_{i=1}^n X_{\gamma_{2 i-1}}^{k_{2 i-1}} \prod_{i=1}^n X_{\gamma_{2 i}}^{k_{2 i}}~.     
\end{eqsp}
\end{widetext}
To compute the trace of the operator, we use the quantum torus algebra to commute the $X_{-\gamma_{2 i-1}}^{\ell_{2 i-1}}$ past $X_{-\gamma_{2 i}}^{\ell_{2 i}}$ and then use the trace formula \eqref{eq:trace_X} noting that the theory has no flavor charges. 
This gives
\begin{eqsp}\label{eq:A2n_mac_conj}
    &\CI^{A_{2n}}_{\text{M}}(q,T)=(q)_{\infty}^n(T q)_{\infty}^n \\&\times\sum_{k_1, \ldots, k_{2 n} \geq 0}  \frac{T^{\sum_{i=1}^nk_{2i-1}}q^{\sum_{i=1}^nk_{2i}+\sum_{i=1}^{2n-1}k_{i}k_{i+1}}}{\prod_{i=1}^n(q)_{k_{2 i-1}}^2 \prod_{i=1}^n(q)_{k_{2 i}}(T q)_{k_{2 i}}}.
\end{eqsp}
For first few values of $n$, one can compute the series and find perfect match with the known form of the Macdonald index \cite{Bhargava:2023hsc,Andrews:2025krn,Song:2016yfd,Song:2015wta}. In fact, we can go further and prove that the above series is identical to the known form of the Macdonald index: 
\begin{eqsp}\label{eq:A_2n_mac_known}
    \CI^{A_{2n}}_{\text{M}}(q,T)=\sum_{\ell_1, \ldots, \ell_n \geq 0} \frac{T^{\ell_1+\ldots+\ell_n} q^{\ell_1^2+\ldots+\ell_n^2+\ell_1+\ldots+\ell_n}}{(q)_{\ell_1-\ell_2}(q)_{\ell_2-\ell_3} \ldots(q)_{\ell_{n-1} \ell_n}(q)_{\ell_n}}~ .
\end{eqsp}
The proof of this nontrivial identity is presented in Appendix \ref{app:proof}.

The BPS quiver of the $(A_1,A_{2n+1})$ theory in the source/sink chamber is shown in Figure \ref{fig:A2n+1}.
The computation of the trace for $(A_1,A_{2n+1})$ is similar to the $(A_1,A_{2n})$ case albeit with the complication of a flavor generator. There is U(1) flavor symmetry for $n>1$ and SU(2) flavor symmetry for $n=1$. Following the calculation in \cite{Cordova:2015nma}, we obtain
\begin{widetext}
\begin{eqsp}
\CI^{A_{2n+1}}_{\text{M}}(q,T,z)=(q)_{\infty}^n(T q)_{\infty}^n \sum_{\substack{\ell_1, \cdots, \ell_{2 n+1}\geq 0\\ k_1, \cdots, k_{2 n+1}\geq 0}}&\frac{(-1)^{\sum_{i=1}^{2 n+1}\left(k_i+\ell_i\right)} T^{\frac{1}{2} \sum_{i=1}^{n+1}\left(k_{2i-1}+\ell_{2i-1}\right)}q^{\frac{1}{2} \sum_{i=1}^{2 n+1}\left(k_i+\ell_i\right)+\sum_{j=1}^n \ell_{2 j}\left(\ell_{2 j-1}+\ell_{2 j+1}\right)}}{\prod_{i=1}^{n+1}(q)_{k_{2i-1}}(q)_{\ell_{2i-1}}\prod_{i=1}^{n}(Tq)_{k_{2i}}(q)_{\ell_{2i}}}
\\&\times\mathrm{Tr}[X_{\gamma_f}]^{\ell_1-k_1}\prod_{i=1}^n \delta_{k_{2 i}, \ell_{2 i}} \prod_{j=1}^n \delta_{(-1)^{j+1} k_1+k_{2 j+1},(-1)^{j+1} \ell_1+\ell_{2 j+1}} ,
\end{eqsp}
\end{widetext}
where we take $\mathrm{Tr}[X_{\gamma_f}]=z^2$ for $n=1$ and $\mathrm{Tr}[X_{\gamma_f}]=(-1)^{(n+1)}z$ for $n>1$ \cite{Cordova:2015nma}. 
Let $\chi_{\bm{n}}(z)$ denote the character of the $n$-dimensional representation of SU(2). The first few terms in the expansion are shown below.\newpage
\begin{eqsp}
&\CI_{\text{M}}^{A_3}(q,T,z)
=1+qT\chi_{\mathbf{3}}+q^2T\left(\chi_{\mathbf{1}}+\chi_{\mathbf{3}}\right)+q^2T^2\chi_{\mathbf{5}} \\&+q^3T\left(\chi_{\mathbf{1}}+\chi_{\mathbf{3}}\right)+q^3T^2(\chi_{\bm{3}}+\chi_{\mathbf{5}})+q^3 T^3\chi_{\mathbf{7}}\\&+q^4T(\chi_{\mathbf{1}}+\chi_{\mathbf{3}}) 
+q^4T^2(\chi_{\mathbf{1}}+2\chi_{\mathbf{3}}+2\chi_{\bm{5}})\\&+q^4T^3(\chi_{\mathbf{5}}+\chi_{\mathbf{7}})+q^4T^4\chi_{\bm{9}}+\dots
\\&\CI_{\text{M}}^{A_5}(q,T,z)
=1+qT+q^{\frac{3}{2}}T^{\frac{3}{2}}(z+z^{-1})+q^2(2T+T^2)\\&+q^{\frac{5}{2}}(T^{\frac{3}{2}}+T^{\frac{5}{2}})(z+z^{-1})+2q^3(T+T^2)\\&+q^3T^3(1+z^{-2}+z^2)+q^{\frac{7}{2}}(T^{\frac{3}{2}}+2T^{\frac{5}{2}}+T^{\frac{7}{2}})(z+z^{-1})\\&+q^4(2T+5T^2+T^3+(1+z^{2}+z^{-2})(T^3+T^4))+\dots
\end{eqsp}
This matches with the results of \cite{Buican:2015tda,Song:2016yfd}.
\subsubsection{$(A_1,D_{k})$ Theory}
\begin{figure}[h]
    \centering
    \includegraphics[width=1\linewidth]{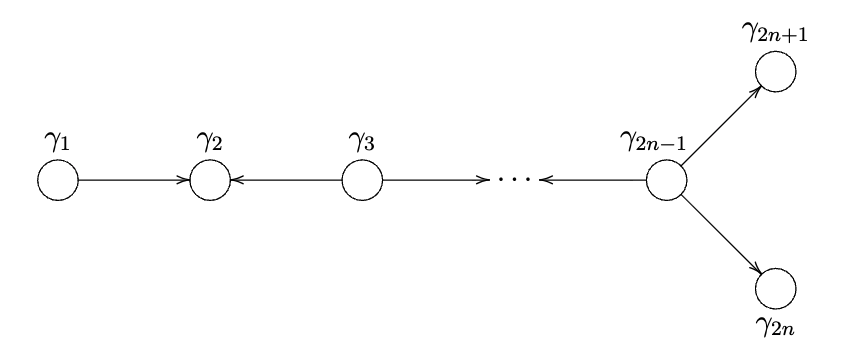}
    \caption{BPS quiver of the $(A_1,D_{2n+1})$ theory in the source/sink chamber. Figure adapted from \cite{Cordova:2015nma}.}
    \label{fig:D2n+1}
\end{figure}

\begin{figure}[h]
    \centering
    \includegraphics[width=1\linewidth]{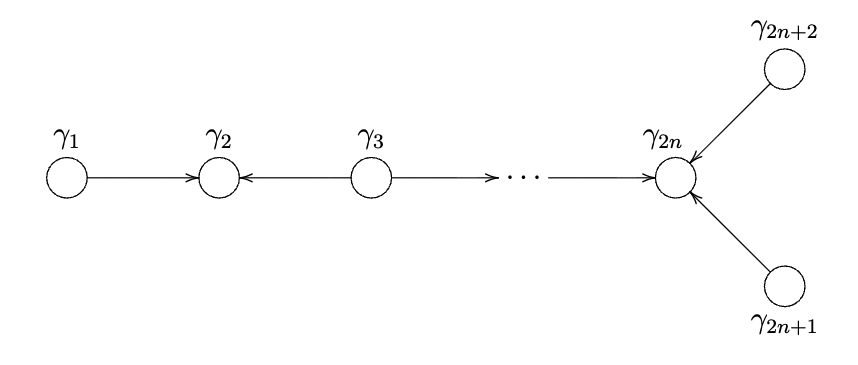}
    \caption{BPS quiver of the $(A_1,D_{2n+2})$ theory in the source/sink chamber. Figure adapted from \cite{Cordova:2015nma}.}
    \label{fig:D2n+2}
\end{figure}

The BPS quiver of the $(A_1,D_{k})$ theory in the source/sink chamber is shown in Figure \ref{fig:D2n+1},\ref{fig:D2n+2}.
Let us start with the $k=2n+1$ case. 
Note that the only node with valency greater than 2 is a source, so the refined KS operator is given by 
\begin{eqsp}
    \CO(q,T)=&\prod_{i=1}^n \widetilde{E}_{q,T}\left(X_{-\gamma_{2 i-1}}\right) \prod_{i=1}^n E_{q,T}\left(X_{-\gamma_{2 i}}\right) E_{q,T}\left(X_{-\gamma_{2 n+1}}\right)\\\times & \prod_{i=1}^n E_{q}\left(X_{\gamma_{2 i-1}}\right) \prod_{i=1}^n E_{q,T}\left(X_{\gamma_{2 i}}\right)E_{q,T}(X_{\gamma_{2n+1}})~.
\end{eqsp}
Following the calculation of trace in \cite{Cordova:2015nma}, we obtain 
\begin{widetext}
\begin{eqsp}
    &\CI_{\text{M}}^{D_{2n+1}}(q,T,z)= (q)_{\infty}^{n}(Tq)_\infty^n \sum_{\substack{\ell_1, \cdots, \ell_{2 n+1}\geq 0\\ k_{2 n}, k_{2 n+1}\geq 0}}\frac{q^{\sum_{i=1}^{2 n+1} \ell_i+\frac{1}{2}\sum_{i,j=1}^{2n+1}b^{D_{2n+1}}_{ij}\ell_i\ell_j}T^{\ell_{2n}+\ell_{2n+1}+\sum_{i=1}^{n-1}\ell_{2i}}z^{2(\ell_{2n+1}-k_{2n+1})}}{(q)_{k_{2 n}}(q)_{k_{2 n+1}}(q)_{\ell_{2 n}}(q)_{\ell_{2 n+1}} \prod_{i=1}^{n-1}(q)_{\ell_{2i}}^2\prod_{i=1}^{n}(q)_{\ell_{2i-1}}(Tq)_{\ell_{2i-1}}} \delta_{k_{2 n}+k_{2 n+1}, \ell_{2 n}+\ell_{2 n+1}}~,
\end{eqsp}
\end{widetext}
where $z$ is the flavor fugacity, $b_{i j}^{D_{2 n+1}}=-C_{i j}^{D_{2 n+1}}+2\delta_{ij}$ and $C_{i j}^{D_{2 n+1}}$ is the Cartan matrix of the $D_{2n+1}$ Lie algebra.
The first few terms for $(A_1,D_5)$ is shown below:
\begin{eqsp}
    &\CI_{\text{M}}^{D_5}(q,T,z)=1+qT\chi_{\bm{3}}+q^2T(\chi_{\bm{1}}+\chi_{\bm{3}})+q^2T^2\chi_{\bm{5}}\\&+q^3T(\chi_{\bm{1}}+\chi_{\bm{3}})+q^3T^2(2\chi_{\bm{3}}+\chi_{\bm{5}})+q^3T^3\chi_{\bm{7}}\\&+q^4T(\chi_{\bm{1}}+\chi_{\bm{3}})+q^4T^2(5\chi_{\bm{3}}+2\chi_{\bm{5}})\\&+q^4T^3(\chi_{\bm{7}}+2\chi_{\bm{5}})+q^4T^4\chi_{\bm{9}}+\dots
\end{eqsp}
The computation for $(A_1,D_{2n+2})$ is similar. We obtain
\begin{widetext}
\begin{eqsp}
\begin{aligned}
\mathcal{I}^{D_{2 n+2}}_{\text{M}}(q,T, x, y)  &=(q)_{\infty}^{n}(Tq)_\infty^n \sum_{\substack{\ell_1, \cdots, \ell_{2 n+2}\ge 0\\ k_1, \cdots, k_{2 n+2}\geq0}}\frac{(-1)^{\sum_{i=1}^{2 n+2}\left(k_i+\ell_i\right)} q^{\frac{1}{2} \sum_{i=1}^{2 n+2}\left(k_i+\ell_i\right)+\frac{1}{2} \sum_{i, j=1}^{2 n+2} b_{i j}^{D_{2 n+2}} \ell_i \ell_j}}{(q)_{k_{2n+1}}(q)_{\ell_{2n+1}}(q)_{k_{2n+2}}(q)_{\ell_{2n+2}}\prod_{i=1}^{n}(Tq)_{k_{2i}}(q)_{\ell_{2i}}\prod_{i=1}^{n}(q)_{k_{2i-1}}(q)_{\ell_{2i-1}}} \\
& \times T^{\frac{1}{2}(k_{2n+1}+k_{2n+2}+\ell_{2n+1}+\ell_{2n+2})+\frac{1}{2}\sum_{i=1}^{n}(k_{2i-1}+\ell_{2i-1})}\left(\operatorname{Tr}\left[X_{\gamma_{f_1}}\right]\right)^{\ell_{2 n+2}-k_{2 n+2}}\left(\operatorname{Tr}\left[X_{\gamma_{f_2}}\right]\right)^{\ell_1-k_1}
\\&\times\left(\prod_{i=1}^n \delta_{k_{2 i}, \ell_{2 i}}\right) \left(\prod_{i=1}^{n-1} \delta_{(-1)^{i+1} k_1+k_{2 i+1},(-1)^{i+1} \ell_1+\ell_{2 i+1}}\right) \delta_{(-1)^{n+1} k_1+k_{2 n+1}+k_{2 n+2},(-1)^{n+1} \ell_1+\ell_{2 n+1}+\ell_{2 n+2}},
\end{aligned}
\end{eqsp}
where $b_{i j}^{D_{2 n+2}}=-C_{i j}^{D_{2 n+2}}+2\delta_{ij}$ and $C_{i j}^{D_{2 n+2}}$ is the Cartan matrix of the $D_{2n+2}$ Lie algebra and we follow the normalization for $\mathrm{SU}(2)\times\mathrm{U(1)}$ flavor fugacities given in \cite{Cordova:2015nma}:
\begin{eqsp}
\operatorname{Tr}\left[X_{\gamma_{f_1}}\right]=y^2, \quad \operatorname{Tr}\left[X_{\gamma_{f_2}}\right]= \begin{cases}x y & \text { if } n \text { is odd~, } \\ -\frac{x}{y} & \text { if } n \text { is even~. }\end{cases}
\end{eqsp}
\end{widetext}
For $(A_1,D_4)$, the flavor symmetry is enhanced to $\mathrm{SU(3)}$ and the fugacities $x,y$ are related to the SU(3) flavor fugacities as $x=z_2^{-3/2},y=z_1z_2^{-1/2}$ \cite{Cordova:2015nma}. Let $\chi_{\bm{n}},\widetilde{\chi}_{\bm{n}}$ denote the character of the $n$-dimensional representation of SU(2),SU(3) respectively. Then the first terms in the expansion of the Macdonald index is given by
\begin{widetext}
\begin{eqsp}
&\mathcal{I}^{D_{4}}_{\text{M}}(q,T, z_1,z_2)=1+qT\widetilde{\chi}_{\bm{8}}+q^2T(\widetilde{\chi}_{\bm{1}}+\widetilde{\chi}_{\bm{8}})+q^2T^2\widetilde{\chi}_{\bm{27}}+q^3T(\widetilde{\chi}_{\bm{1}}+\widetilde{\chi}_{\bm{8}})+q^3T^2(\widetilde{\chi}_{\bm{8}}+\widetilde{\chi}_{\bm{10}}+\widetilde{\chi}_{\overline{\bm{10}}}+\widetilde{\chi}_{\bm{27}})+q^3T^3\widetilde{\chi}_{\bm{64}}+\dots
\\&\mathcal{I}^{D_{6}}_{\text{M}}(q,T, x,y)=1+qT(\chi_{\bm{1}}+\chi_{\bm{3}})+q^{\frac{3}{2}}T^{\frac{3}{2}}(x+x^{-1})\chi_{\bm{2}}+q^2T(2\chi_{\bm{1}}+\chi_{\bm{3}})+q^2T^2(\chi_{\bm{1}}+\chi_{\bm{3}}+\chi_{\bm{5}})+q^{\frac{5}{2}}T^{\frac{3}{2}}(x+x^{-1})\chi_{\bm{2}}\\&+q^{\frac{5}{2}}T^{\frac{5}{2}}(x+x^{-1})(\chi_{\bm{2}}+\chi_{\bm{4}})+q^3T(2\chi_{\bm{1}}+\chi_{\bm{3}})+q^3T^2(2\chi_{\bm{1}}+4\chi_{\bm{3}}+\chi_{\bm{5}})+q^3T^3\left((\chi_{\bm{1}}+(x^{-2}+x^{2}+1)\chi_{\bm{3}}+\chi_{\bm{5}}+\chi_{\bm{7}})\right)+\dots
\end{eqsp}    
\end{widetext}
The Macdonald index for $(A_1,D_4)$ matches with \cite[Table 2]{Buican:2015tda} and $(A_1,D_6)$ matches with the general formula of \cite{Buican:2015tda}. 
\subsubsection{$(A_1,E_n)$ Theory}
\begin{figure}[h]
    \centering
    \includegraphics[width=0.85\linewidth]{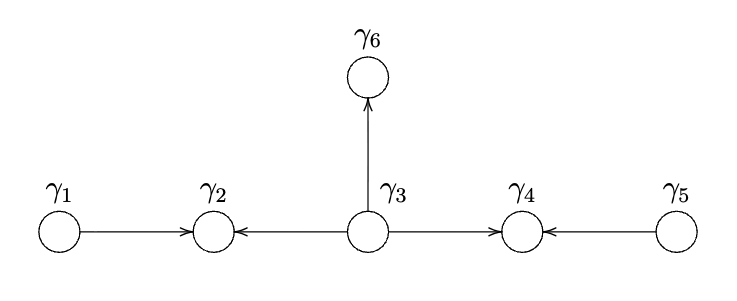}
    \caption{BPS quiver of the $(A_1,E_6)$ theory in the source/sink chamber. Figure adapted from \cite{Cordova:2015nma}.}
    \label{fig:DE6}
\end{figure}

Let us start with $(A_1,E_6)$ theory. 
The BPS quiver for the $(A_1,E_6)$ theory is shown in Figure \ref{fig:DE6}.
\begin{figure}[h]
    \centering
    \includegraphics[width=1\linewidth]{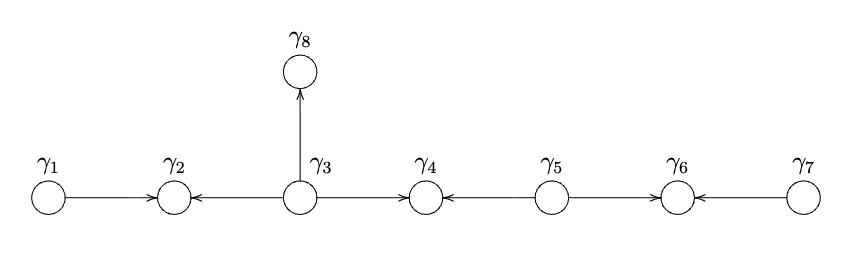}
    \caption{BPS quiver of the $(A_1,E_8)$ theory in the source/sink chamber. Figure adapted from \cite{Cordova:2015nma}.}
    \label{fig:DE8}
\end{figure} 
The refined KS operator is given by 
\begin{eqsp}
\CO(q,T)=&\prod_{i=1}^{3}\widetilde{E}_{q,T}(X_{-\gamma_{2i-1}})\prod_{i=1}^{3}E_{q,T}(X_{-\gamma_{2i}})\\\times&\prod_{i=1}^{3}E_{q}(X_{\gamma_{2i-1}})\prod_{i=1}^{3}E_{q,T}(X_{\gamma_{2i}})~.
\end{eqsp}
Computing the trace, the Macdonald index is given by 
\begin{eqsp}
&\mathcal{I}^{E_6}_{\text{M}}(q,T)\\&=(q)_{\infty}^3(qT)_\infty^3 \sum_{\ell_1, \cdots, \ell_6=0}^{\infty} \frac{T^{\ell_2+\ell_4+\ell_6}q^{\sum_{i=1}^6 \ell_i+\frac{1}{2} \sum_{i, j=1}^6 b_{i j}^{E_6} \ell_i \ell_j}}{\prod\limits_{i=1}^{3}(qT)_{\ell_{2i-1}}(q)_{\ell_{2i-1}}\prod\limits_{i=1}^3(q)_{\ell_{2i}}^2}~
\\&=1+q^2T+q^3(T+T^2)+q^4(T+2T^2)+q^5 (T+2 T^2)\\&+q^6\left(T+3 T^2+2 T^3\right)+\dots
\end{eqsp}
where $b_{i j}^{E_6}=-C_{i j}^{E_6}+2\delta_{ij}$ and $C_{i j}^{E_6}$ is the Cartan matrix of the $E_6$ Lie algebra. The BPS quiver of $(A_1,E_8)$ theory is shown in Figure \ref{fig:DE8}. 
We now describe the computation of the Macdonald index for the $(A_1,E_7)$ theory. Unlike the other $(A_1,E_n)$ theories, $(A_1,E_7)$ has a U(1) flavor symmetry. Note also that there is a source node with valency greater than 2. Thus we must use \eqref{eq:KS_source>2} to write the refined KS operator. Performing the computation parallel to \cite{Cordova:2015nma} with the refined KS operator, we get
\begin{widetext}
\begin{eqsp}
\mathcal{I}_{\text{M}}^{E_7}(q,T, z) &=(q)_{\infty}^3 (qT)_{\infty}^3 \sum_{\substack{\ell_1, \cdots, \ell_7\geq 0\\ k_1, \cdots, k_7\geq 0}}\frac{(-1)^{\sum_{i=1}^7\left(k_i+\ell_i\right)} T^{\frac{1}{2}(k_7+\ell_7)+\frac{1}{2}\sum_{i=1}^3(k_{2i}+\ell_{2i})}q^{\frac{1}{2} \sum_{i=1}^7\left(k_i+\ell_i\right)} q^{k_2\left(\ell_1+\ell_3\right)+k_4\left(\ell_3+\ell_5\right)+k_6 \ell_5+k_7 \ell_3}}{(q)_{k_7}(q)_{\ell_7}\prod_{i=1}^3(qT)_{k_{2i-1}}(q)_{\ell_{2i-1}}\prod_{i=1}^3(q)_{k_{2i}}(q)_{\ell_{2i}}}\\&\hspace{4cm}\times (-z)^{-\ell_7+k_7} \delta_{k_4+k_7, \ell_4+\ell_7} \delta_{k_6-k_7, \ell_6-\ell_7} \prod_{i \in\{1,2,3,5\}} \delta_{k_i, \ell_i}
\\&=1+qT+q^{\frac{3}{2}}T^{\frac{3}{2}}(z+z^{-1})+q^2(2T+T^2)+q^{\frac{5}{2}}(T^{\frac{3}{2}}+T^{\frac{5}{2}})(z+z^{-1})\\&+q^3(2T+3T^2+T^3(z^{-2}+1+z^2))+q^{\frac{7}{2}}(T^{\frac{3}{2}}+3T^{\frac{3}{2}}+T^{\frac{7}{2}})(z+z^{-1})+\dots,
\end{eqsp}    
\end{widetext}
where $z$ is the flavor fugacity. \\
The computation of the Macdonald index of $(A_1,E_8)$ theory identical to that of $(A_1,E_6)$ theory and is given by 
\begin{eqsp}
&\mathcal{I}^{E_8}_{\text{M}}(q,T)\\&=(q)_{\infty}^4(qT)_\infty^4 \sum_{\ell_1, \cdots, \ell_8=0}^{\infty} \frac{T^{\sum\limits_{i=1}^4\ell_{2i}}q^{\sum\limits_{i=1}^8 \ell_i+\frac{1}{2} \sum\limits_{i, j=1}^8 b_{i j}^{E_8} \ell_i \ell_j}}{\prod\limits_{i=1}^{4}(qT)_{\ell_{2i-1}}(q)_{\ell_{2i-1}}\prod\limits_{i=1}^4(q)_{\ell_{2i}}^2}
\\&=1+q^2T+q^3(T+T^2)+q^4(T+2T^2)\\&+q^5(T+2T^2+T^3)+q^6(T+3T^2+3T^3)+\dots,
\end{eqsp}
where $b_{i j}^{E_8}=-C_{i j}^{E_8}+2\delta_{ij}$ and $C_{i j}^{E_8}$ is the Cartan matrix of the $E_8$ Lie algebra. It is expected that $(A_1,E_{6,8})\cong (A_2,A_{3,4})$ \cite{Wang:2015mra}. We confirmed that the Macdonald indices of $(A_1,E_{6,8})$ theory matches with the Macdonald indices of $(A_2,A_{3,4})$  theory computed in \cite{Watanabe:2019ssf,Agarwal:2018zqi}.

\section{Discussion}
The intriguing relation we have found between Macdonald index of a 4d $\CN=2$ SCFT and the trace of the modified KS operator raises a lot of interesting questions. The modified function $E_{q,T}(z)$ can be recognized as the Macdonald index of a half-hypermultiplet. This means that our prescription assigns a full hypermultiplet to the source nodes in case the BPS quiver does not have a source node with valency greater than 2 and to the sink nodes in case the BPS quiver has a source node with valency greater than 2. But the function $\widetilde{E}_{q,T}$ does not seem to be related to any superconformal multiplet. Perhaps a simplified model of BPS bound states in terms of ``halos'' \cite{Denef:2007vg,Gaiotto:2010be,Gaiotto:2010okc} could be useful in understanding the physical meaning of the function $\widetilde{E}_{q,T}$. Using bound state halos in terms of noninteracting bosons and fermions, \cite{Denef:2007vg} could prove some special cases of the KS wall crossing formula. Hence, it will be useful to understand (and possibly prove) the Cordova-Shao conjecture and our prescription in terms of halos. 
\par We list here some possible future directions. (1) The first order problem is to perform more checks of our conjecture. (2) Secondly, we would like to understand how the refined KS operator transforms under quiver mutations \cite{Cecotti:2010fi,Alim:2011ae}. More generally,  we would like to give a chamber independent prescription for the refined KS operator. Once we have a chamber independent prescription, our formulae for Argyres-Douglas theories give strong evidence that the modified KS operator is a wall-crossing invariant.
(3) Finally, we would like to interpret our prescription (and in particular the Cordova-Shao) conjecture in terms of BPS halos. 
\\\\
\textbf{Acknowledgments.} The authors would like to thank Gregory Moore, Leonardo Rastelli and Anirudh Deb for comments on the draft and Ashoke Sen for some useful discussions on related topics.  R.K.S. would like to thank LITP, Stanford University for hospitality where this work began.  The work of A.B. is supported by the US Department of Energy under grant DE-SC1019775. The work of R.K.S. is supported by the US
Department of Energy under grant DE-SC0010008.
\bibliographystyle{apsrev4-1}
\bibliography{apssamp}%
\appendix
\onecolumngrid
\section{Proof of a $q$-series identity}\label{app:proof}
In this appendix, we prove that the conjecture \eqref{eq:A2n_mac_conj} for the Macdonald index of $(A_1,A_{2n})$ theory agrees with the known form \eqref{eq:A_2n_mac_known} of the Macdonald index. More precisely, we prove the following
\begin{thm}\label{thm:mac-id-A2n}
We have the following identity
\begin{eqsp}
 (q)_{\infty}^n(T q)_{\infty}^n \sum_{k_1, \ldots, k_{2 n} \geq 0} T^{k_1+k_3+\ldots+k_{2 n-1}} q^{k_1+\ldots+k_{2 n}} &\frac{q^{k_1 k_2+k_2 k_3+\ldots+k_{2 n-1} k_{2 n}}}{\prod_{i=1}^n(q)_{k_{2 i-1}}^2 \prod_{i=1}^n(q)_{k_{2 i}}(T q)_{k_{2 i}}} \\
& =\sum_{\ell_1, \ldots, \ell_n \geq 0} \frac{T^{\ell_1+\ldots+\ell_n} q^{\ell_1^2+\ldots+\ell_n^2+\ell_1+\ldots+\ell_n}}{(q)_{\ell_1-\ell_2}(q)_{\ell_2-\ell_3} \ldots(q)_{\ell_{n-1}- \ell_n}(q)_{\ell_n}} .
\end{eqsp}
\end{thm}
We will prove this in three steps. 
Let us denote the LHS by $L$ and RHS by $R$. For $1\leq i\leq n-1$, define
\begin{equation}
\begin{split}
S_i=(q)^{n-i}(T q)^{n-i}  \sum_{\substack{k_1, \ldots, k_{2 n-2 i} \geq 0 \\
n_1, \ldots, n_i \geq 0}} &T^{k_1+\ldots+k_{2 n-2 i}} q^{k_1+\ldots+k_{2 n-2 i}} \frac{q^{k_1 k_2+\ldots+k_{2 n-2 i-1} k_{2 n-2 i}}}{\prod_{r=1}^{n-i}(q)_{k_{2 r-1}}^2 \prod_{r=1}^{n-i}(q)_{k_{2 r}}(T q)_{k_{2 r}}} \\
& \times \frac{q^{n_1^2+\ldots+n_i^2+n_1+\ldots+n_i}\left(T q^{n_i+1}\right)_{2 n-2 i}}{(q)_{n_1-n_2}(q)_{n_2-n_3} \ldots(q)_{n_{i-1}-n_i}(q)_{n_i}}~ .
\end{split}
\end{equation}
We prove the result in three steps.
\begin{enumerate}
    \item $L = S_1$.

\item  $R = S_{n-1}$.

\item  $S_i = S_{i+1}$ for $1\leq i\leq n-2$.
\end{enumerate}
\noindent We will need the following identities:

\begin{lemma}\label{lemma:int1}
We have
\begin{equation}
\sum_{N \geq 0} \frac{T^N}{(q)_N(c q)_N}=\frac{1}{(t)_{\infty}} \sum_{N \geq 0} \frac{q^{N^2} T^{N^2} c^N}{(q)_N(c q)_N}~.
\end{equation}
\end{lemma}
\textit{Proof.} 
We use \cite[Equation (3.3.13), Page 39]{andrews1998theory}:
\begin{equation}\label{eq:3.3.13_andrews}
\sum_{N=0}^\infty \frac{(a)_N(b)_N t^N}{(q)_N(c)_N} = \frac{(abt/c)_\infty}{(t)_\infty} \sum_{N=0}^\infty \frac{(c/a)_N(c/b)_N(abt/c)^N}{(q)_N(c)_N}~.
\end{equation}
We replace $c$ by $cq$ and take $a,b\to\infty$. Note that
\begin{equation}
\begin{split}
\lim_{a \to b} (c/a)_N a^N = \lim_{a \to 0} (1 - c/a)(1 - c/aq) \cdots (1 - c q^{N-1}/a) a^N \\
= \lim_{a \to 0} (-c)^N \cdot q^{\frac{N(N-1)}{2}}~.
\end{split}
\end{equation}

Thus we get
\begin{equation}
\sum_{N \geq 0} \frac{T^N}{(q)_N(c)_N} = \frac{1}{(t)_\infty} \sum_{N=0}^\infty \frac{c^N q^{N^2-N}}{(q)_N(c)_N} T^N~.
\end{equation}
Replace $ c \to cq $ to get the result. 
\hfill $\square$
\\\\
We also have
\begin{equation}
\sum_{N \geq 0} \frac{T^N}{(q)_N} = \frac{1}{(T,q)_\infty} = \frac{1}{(T)_\infty}
\end{equation}
\begin{lemma}\label{lemma:int2}
We have
\begin{equation}
\sum_{j \geq 0} \frac{\left(c q^{N+1}\right)_j t^j}{(q)_j(c q)_j}=\frac{1}{(t)_{\infty}} \sum_{j=0}^N\binom{N}{j}_q \frac{q^{j^2} c^j t^j}{(c q)_j}~.
\end{equation}
\end{lemma}
\textit{Proof.} 
We get this by taking $ a \to 0 $ and $ b = cq^{N+1} $ in \eqref{eq:3.3.13_andrews}. We have
\begin{equation}
\begin{split}
\sum_{j=0}^{\infty} \frac{\left(c q^{N+1}\right)_j t^j(a)_j}{(q)_j(c q)_j} & =\frac{1}{(t)_{\infty}} \sum_{j=0}^{\infty} \frac{(c q / a)_j a^j\left(c q / c q^{N+1}\right)_j\left(c q^{N+1} t / c q\right)^j}{(q)_j(c q)_j} \\
& =\frac{1}{(t)_{\infty}} \sum_{j=0}^{\infty} \frac{(c q / a)_j a^j\left(q^{-N}\right)_j\left(q^N t\right)^j}{(q)_j(c q)_j}~.
\end{split}
\end{equation}
Now

\begin{equation}
\begin{aligned}
\lim _{a \rightarrow 0}(c q / a) j a^j & =\lim _{a \rightarrow 0}(1-c q / a)\left(1-c q^2 / a\right) \cdots\left(1-c q^{j+1} / a\right) a^j \\
& =(-c)^j q^{1+2+\ldots+j+1}=(-c)^j q^{j(j+1) / 2}~.
\end{aligned}
\end{equation}
Next
\begin{equation}
\left(q^{-N}\right)_j=\left(q^{-N} ; q\right)_j=\frac{(q)_N}{(q)_{N-j}}(-1)^j q^{{j\choose 2}-N j}~.
\end{equation}
We get 
\begin{equation}
\begin{aligned}
\sum_{j=0}^{\infty} \frac{\left(c q^{N+1}\right)_j t^j}{(q)_j(c q)_j} & =\frac{1}{(t)_{\infty}} \sum_{j=0}^{\infty} \frac{c^j q^{\left(j^2+j\right) / 2} q^{j(j-1) / 2-N_j} q^{Nj} T^j}{(q)_j(c q)_j} \frac{(q)_N}{(q)_{N-j}} \\
& =\frac{1}{(T)_{\infty}} \sum_{j=0}^N \frac{c^j q^{j^2} T^j}{(c q)_j}\binom{N}{j}_q .
\end{aligned}
\end{equation}
\hfill $\square$
\\\noindent Introduce the notation  
\begin{align}
H_j &:= \frac{T^{k_1+k_3+\cdots+k_{2j-1}} q^{k_1+k_2
+\cdots+k_{2j}+k_1k_2+\cdots+k_{2j-1} k_{2j}}}{\prod_{r=1}^{j}(q)_{k_{2r-1}}^{2} \prod_{r=1}^{j}(q)_{k_{2r}}(Tq)_{k_{2r}}}
\\
N_j &:= \frac{T^{n_1+\cdots+n_j} q^{n_1^2+\cdots+n_j^2+n_1+\cdots+n_j}}{(q)_{n_1-n_2}(q)_{n_2-n_3}\cdots(q)_{n_{j-1}-n_j}(q)_{n_j}}~.
\end{align}
Then we can write  
\begin{equation}
\begin{split}
H_j = \frac{T^{k_{2j-1}} q^{k_{2j}+k_{2j-1}+k_{2j}k_{2j-1}+k_{2j-1}k_{2j-2}}}{ (q)_{k_{2j-1}}^{2}(q)_{k_{2j}}(tq)_{k_{2j}}}H_{j-1}~.
\end{split}
\end{equation}
and  
\begin{equation}
\begin{split}
N_j=\frac{t^{n_j} q^{n_j^2+n_j}(q)_{n_{j-1}}}{(q)_{n_j}(q)_{n_{j-1}-n_j}} N_{j-1}=t^{n_j} q^{n_j^2+n_j}\binom{n_{j-1}}{n_j}_q N_{j-1}~ .
\end{split}
\end{equation}
We have  
\begin{equation}
S_i = (q)^{n-i}_\infty (tq)^{n-i}_\infty \sum_{\substack{k_{1},\ldots,k_{2n-2i}\ge0 \\ n_1,\ldots,n_i\ge0}} H_{n-i} N_i(Tq^{n_i+1})_{k_{2n-2i}}~.
\end{equation}
Let us first prove that $L=S_1$. 
\begin{prop}
Let $L,S_i$ be as above. Then we have
\begin{eqsp}
    L=S_1~.
\end{eqsp}
\end{prop}
\textit{Proof.} 
We have  
\begin{equation}
\begin{split}
L & =(q)_{\infty}^n(t q)_{\infty}^n \sum_{k_1, \ldots, k_{2 n} \geq 0} H_n=(q)_{\infty}^n(t q)_{\infty}^n \sum_{k_1, \ldots, k_{2 n} \geq 0} \frac{t^{k_{2 n-1}} q^{k_{2 n}+k_{2 n-1}+k_{2 n} k_{2 n-1}+k_{2 n-1} k_{2 n-2}}}{(q)_{k_{2 n-1}}^2(q)_{k_{2 n}}(t q)_{k_{2 n}}} H_{n-1} \\
& =(q)_{\infty}^n(t q)_{\infty}^n \sum_{k_{2 n} \geq 0} \frac{q^{k_{2 n}+k_{2 n} k_{2 n-1}}}{(q)_{k_{2 n}}(t q)_{k_{2 n}}} \sum_{k_1, \ldots, k_{2 n-1} \geq 0} \frac{t^{k_{2 n-1}} q^{k_{2 n-1}+k_{2 n-1} k_{2 n-2}}}{(q)_{k_{2 n-1}}^2} H_{n-1}~.
\end{split}
\end{equation}
Using Lemma \ref{lemma:int1} with $T=q^{1+k_{2n-1}}, c=T$, see get  
\begin{equation}
\sum_{k_{2n}\ge0} \frac{q^{(1+k_{2n-1})k_{2n}}}{(q)_{k_{2n}}(Tq)_{k_{2n}}} = \frac{1}{(q^{1+k_{2n-1}})_\infty} \sum_{n_1\ge0} \frac{q^{n_1^2} q^{n_1(1+k_{2n-1})}T^{n^1}}{(q)_{n_1}(Tq)_{n_1}}~.
\end{equation}
Substituting this we get
\begin{equation}
\begin{aligned}
L & =(q)_{\infty}^n(t q)_{\infty}^n \sum_{k_1, \ldots, k_{2 n-1} \geq 0} \frac{t^{k_{2 n-1}} q^{k_{2 n-1}+k_{2 n-1} k_{2 n-2}}}{(q)_{k_{2 n-1}}^2} H_{n-1} \frac{q^{n_1^2} q^{n_1\left(1+k_{2 n-1}\right)} t^{n_1}}{\left(q^{1+k_{2 n-1}}\right)_{\infty}(q)_{n_1}(t q)_{n_1}} \\
& =(q)_{\infty}^n(t q)_{\infty}^n \sum_{\substack{k_1, \ldots, k_{2 n-2} \geq 0 \\
n_1 \geq 0}} H_{n-1} \frac{q^{n_1^2} t^{n_1}}{(q)_{n_1}(t q)_{n_1}} \sum_{k_{2 n-1} \geq 0} \frac{t^{k_{2 n-1}} q^{k_{2 n-1}+k_{2 n-1} k_{2 n-2}}}{(q)_{k_{2 n-1}}^2} \frac{q^{n_1\left(1+k_{2 n-1}\right)}}{\left(q^{1+k_{2 n-1}}\right)_{\infty}} \\
& =(q)_{\infty}^n(t q)_{\infty}^n \sum_{\substack{k_1, \ldots, k_{2 n-2} \geq 0 \\
n_1 \geq 0}} H_{n-1} \frac{q^{n_1^2+n_1} t^{n_1}}{(q)_{n_1}(t q)_{n_1}} \sum_{k_{2 n-1} \geq 0} \frac{t^{k_{2 n-1}} q^{k_{2 n-1}\left(1+k_{2 n-2}+n_1\right)}\left(q^{1+k_{2 n-1}}\right)_{\infty}}{(q)_{k_{2 n-1}}\left(q^{1+k_{2 n-1}}\right)_{\infty}(q)_{\infty}} \\
& =(q)_{\infty}^{n-1}(t q)_{\infty}^n \sum_{\substack{k_1, \ldots, k_{2 n-2} \geq 0 \\
n_1 \geq 0}} H_{n-1} \frac{q^{n_1^2+n_1} t^{n_1}}{(q)_{n_1}(t q)_{n_1}} \sum_{k_{2 n-1} \geq 0} \frac{t^{k_{2 n-1}} q^{k_{2 n-1}\left(1+k_{2 n-2}+n_1\right)}}{(q)_{k_{2 n-1}}} \\
& =(q)_{\infty}^{n-1}(t q)_{\infty}^n \sum_{k_1, \ldots, k_{2 n-2} \geq 0} H_{n-1} \frac{q^{n_1^2+n_1} t^{n_1}}{(q)_{n_1}(t q)_{n_1}} \frac{1}{\left(t q^{n+k_{2 n-2}+1}\right)_{\infty}}~,
\end{aligned}
\end{equation}
where we used
\begin{eqsp}
&\sum_{k_{2 n-1} \geq 0} \frac{t^{k_{2 n-1}} q^{k_{2 n-1}\left(1+k_{2 n-2}+n_1\right)}}{(q)_{k_{2 n-1}}}=\sum_{k_{2 n-1} \geq 0} \frac{\left(t q^{1+n_1+k_{2 n-2}}\right)^{k_{2 n-1}}}{(q)_{k_{2 n-1}}}=\frac{1}{\left(t q^{1+n_1+k_{2 n-2}}\right)_{\infty}} ~.   
\end{eqsp}
Thus we have
\begin{equation}
\begin{aligned}
L & =(q)_{\infty}^{n-1}(t q)_{\infty}^n \sum_{\substack{k_{1},\dots, k_{2 n-2}\geq 0 \\
n_1 \geq 0}} H_{n-1} N_1 \cdot \frac{\left(t q^{n_1+1}\right)_{\infty}}{(t q)_{\infty}\left(t q^{1+n_1+k_{2 n-2}}\right)_{\infty}} \\
& =(q)_{\infty}^{n-1}(t q)_{\infty}^{n-1} \sum_{\substack{k_1, \ldots, k_{2 n-2} \geq 0 \\
n_1 \geq 0}} H_{n-1} N_1\left(t q^{n_1+1}\right)_{k_{2 n-2}} \\
& =S_1 ~.
\end{aligned}
\end{equation}
\hfill $\square$\\\\
We now prove that
\begin{equation}
S_{n-1} = R.
\end{equation}
\begin{prop}
Let $R,S_i$ be as above. Then we have 
\begin{eqsp}
    R=S_{n-1}~.
\end{eqsp}
\end{prop}
\textit{Proof.} 
We have
\begin{equation}
\begin{split}
S_{n-1} &= (q)_\infty(tq)_\infty\sum_{n_1,\ldots,n_{n-1}\geq 0}N_{n-1}\frac{T^{k_1} q^{k_1+k_2+k_1 k_2}}{(q)_{k_1}^2(q)_{k_2}(tq)_{k_2}}(tq^{n_{n-1}+1})_{k_2} \\
&= (q)_\infty(tq)_\infty\sum_{n_1,\ldots,n_{n-1}\geq 0}N_{n-1}\frac{T^{k_1} q^{k_1}}{(q)_{k_1}^2}\sum_{k_2\geq 0}\frac{q^{k_2+k_1 k_2}}{(q)_{k_2}(tq)_{k_2}}(tq^{n_{n-1}+1})_{k_2}
\end{split}
\end{equation}
Using Lemma \ref{lemma:int2} for the sum over $k_2$ with $c=T$, $T=q^{k_1+1}$, we get
\begin{equation}
\begin{split}
\sum_{k_2\geq 0}\frac{q^{k_2+k_1 k_2}}{(q)_{k_2}(tq)_{k_2}}(tq^{n_{n-1}+1})_{k_2} &= \sum_{k_2\geq 0}\frac{q^{(1+k_1)k_2}(tq^{n_{n-1}+1})_{k_2}}{(q)_{k_2}(tq)_{k_2}} \\
&= \frac{1}{(q^{1+k_2})_{\infty}}\sum_{n_1,\ldots,n_{n-1}\geq 0}^{n_{n-1}} \binom{n_{n-1}}{n_n}_q \frac{q^{n_n^2}T^{n_n} q^{n_n(1+k_1)}}{(Tq)_{n_n}}~.
\end{split}
\end{equation}
Substituting we get
\begin{equation}
\begin{split}
S_{n-1} &= (q)_\infty(tq)_\infty\sum_{\substack{n_1,\ldots,n_{n-1}\geq 0\\k_1\geq 0}}N_{n-1}\frac{T^{k_1} q^{k_1}}{(q)_{k_1}^2(q^{1+k_2})_{\infty}}\sum_{n_1,\ldots,n_{n-1}\geq 0}^{n_{n-1}} \binom{n_{n-1}}{n_n}_q \frac{q^{n_1^2+n_n} q^{n_nk_1}T^{n_n}}{(q)_{n_n}(Tq)_{n_n}} \\
&= (q)_\infty(tq)_\infty\sum_{n_1,\ldots,n_{n-1}\geq 0}N_{n-1}\binom{n_{n-1}}{n_n} \frac{q^{n_1^2+n_n}T^{n_n}}{(q)_{n_n}(Tq)_{n_n}}\sum_{k_1\geq 0}\frac{T^{k_1} q^{k_1+n_n k_1}}{(q^{k_1+1})_{\infty}(q)_{k_1}^{2}}~.
\end{split}
\end{equation}
We have
\begin{equation}
\sum_{k_1\geq 0}\frac{T^{k_1} q^{k_1+n_n k_1}}{(q^{k_1+1})_{\infty}(q)_{k_1}^{2}} = \sum_{k_2\geq 0}\frac{T^{k_1} q^{(1+n_n)k_1}}{(q)_{k_1}(q)_{\infty}} = \frac{1}{(tq^{1+n_n})_{\infty}}\frac{1}{(q)_{\infty}}~.
\end{equation}
Thus we get
\begin{equation}
\begin{split}
S_{n-1} &= (tq)_\infty \sum_{n_1, \ldots, n_n \geq 0} N_{n-1} \binom{n_{n-1}}{n_n} \frac{q^{n_1^2+n_n}T^{n_n}}{(q)_{n_n}(Tq)_{n_n}} \frac{1}{ (tq^{n_n+1})_\infty} \\
&= \sum_{n_1, \ldots, n_n \geq 0} N_{n-1} \binom{n_{n-1}}{n_n}_q q^{n_1^2+n_n}T^{n_n} \\
&= \sum_{n_1, \ldots, n_n \geq 0} N_n \\&= R~.
\end{split}
\end{equation}
\hfill $\square$\\\\
We finally prove that $S_i=S_{i+1}$.
\begin{prop}
Let $S_i$ be as above. Then we have  
\begin{equation}
S_i = S_{i+1} \quad \text{for} \quad 1 \leq i \leq n-2~.
\end{equation}
\end{prop}
\textit{Proof.} 
For ease of notation, we will write
\begin{equation}
j_2 = k_{2n-2i}, \quad j_1 = k_{2n-2i-1}~.
\end{equation}
Then we have
\begin{equation}
\begin{split}
S_i &= (q)^{n-i}_\infty (tq)^{n-i}_\infty \sum_{\substack{k_{1},\cdots,k_{2n-2i-2} \geq 0\\j_1,j_2\geq 0\\n_1,\dots,n_i\geq 0}} H_{n-i-1} \frac{T^{j_1} q^{j_2 + j_1 + j_2 j_1 + j_1k_{2n-2i-2}} }{(q)_{j_1}^2 (q)_{j_2} (tq)_{j_2}} \cdot N_i (tq^{n_i+1})_{j_2} \\
&= (q)^{n-i}_\infty (tq)^{n-i}_\infty \sum_{\substack{k_{1},\cdots,k_{2n-2i-2} \geq 0\\j_1\geq 0\\n_1,\dots,n_i\geq 0}} H_{n-i-1} N_i \frac{T^{j_1} q^{j_1 + j_1k_{2n-2i-2}} }{(q)_{j_1}^2 } \sum_{j_2 \geq 0} \frac{q^{j_2 + j_1 j_2}}{(q)_{j_2} (tq)_{j_2}} (tq^{n_i+1})_{j_2} \\
&= (q)^{n-i}_\infty (tq)^{n-i}_\infty \sum_{\substack{k_{1},\cdots,k_{2n-2i-2} \geq 0\\j_1\geq 0\\n_1,\dots,n_i,n_{i+1}\geq 0}} H_{n-i-1} N_i \frac{T^{j_1} q^{j_1 + j_1k_{2n-2i-2}} }{(q)_{j_1}^2(q^{1+j_1})_\infty}{n_i\choose n_{i+1}}_q \frac{q^{n_{i+1}^2} T^{n_{i+1}} q^{n_{i+1} (t+j_1)}}{(tq)_{n_{i+1}}}\\
&= (q)_{\infty}^{n-i-1} (tq)_{\infty}^{n-i} \sum_{\substack{k_{1},\cdots,k_{2n-2i-2} \geq 0\\n_1,\dots,n_i,n_{i+1}\geq 0}} H_{n-i-1} N_{i+1} \frac{1}{(tq^{k_{2n-2i-2}+n_{i+1}+1})_{\infty} (tq)_{n_{i+1}}} \\
&= (q)_{\infty}^{n-i-1} (tq)_{\infty}^{n-i-1} \sum_{\substack{k_{1},\cdots,k_{2n-2i-2} \geq 0\\n_1,\dots,n_i,n_{i+1}\geq 0}} H_{n-i-1} N_{i+1} (tq^{n_{i+1}+1})_{k_{2n-2i-2}} \\
&= S_{i+1}~.
\end{split}
\end{equation}
\hfill $\square$
\\\\
\noindent\textit{Proof of Theorem \ref{thm:mac-id-A2n}.} Follows from the above propositions and induction on $i$. \hfill $\square$

\end{document}